\documentclass[12pt,preprint]{aastex}    

\slugcomment{to appear in Astrophysical Journal August 10 2006}

\begin{document}

\shorttitle{Astrometry and Photometry with Coronagraphs}
\shortauthors{Sivaramakrishnan \& Oppenheimer}


\newcommand \cf     {cf.}
\newcommand \eg     {{\it e.g., }}
\newcommand \etc    {{\it etc}}
\newcommand \etal   {{\it et~al.}}
\newcommand \ie     {{\it i.e.,}}
\newcommand \viz    {{\it viz.,}}
\newcommand \lap    {$\stackrel{<}{\sim}$}
\newcommand \gap    {$\stackrel{>}{\sim}$}
\newcommand \fwhm   {${FWHM}$}
\newcommand \both   {(\Phi + {\Psi}_k)}
\newcommand \pk     {{\Psi}_k}
\newcommand \ph     {\Phi}
\newcommand \Sumk   {\Sigma_k\ }
\newcommand \eq     {\,=\,}                 

\newcommand \sgn   {{\rm{sgn}}}
\newcommand \jinc   {{\rm{jinc}}}
\newcommand \conv {{ * }}

\newcommand \Dlam 	{D_{\lambda}}
\newcommand \dlam 	{d_{\lambda}}
\newcommand \blam 	{b_{\lambda}}
\newcommand \ssgn 	{\,\, ^2\rm sgn}			
\newcommand \shah 	{{\rm III}}     			
\newcommand \sshah 	{{\,\, ^2\rm III}}			
\newcommand \ppi 	{\,\, ^2\Pi}				
\newcommand \sinc 	{\,\,  {\rm sinc}}		    
\newcommand \ssinc 	{\,\, ^2 {\rm sinc}}		
\newcommand \bx 	{{\bf x}}
\newcommand \bk 	{{\bf k}}
\newcommand \e 		{{\epsilon}}
\newcommand \g 		{{g}}

\title{Astrometry and Photometry with Coronagraphs}

\author{Anand Sivaramakrishnan\altaffilmark{1,2} \& Ben R. Oppenheimer\altaffilmark{3} }
	\affil{Department of Astrophysics, 
		American Museum of Natural History, \\
		79th Street at Central Park West, New York, NY 10024}
\altaffiltext{1}{NSF Center for Adaptive Optics}
\altaffiltext{2}{Stony Brook University}
\altaffiltext{3}{Columbia University}
\begin{abstract} 
We propose a solution to the problem of astrometric and photometric
calibration of coronagraphic images with a simple optical device which,
in theory, is easy to use.  Our design uses the Fraunhofer approximation of 
Fourier optics.
Placing a periodic grid of wires (we use a square grid) with
known width and spacing in a pupil plane in front
of the occulting coronagraphic focal plane mask
produces fiducial images of the obscured star at
known locations relative to the star.  
We also derive the intensity of these fiducial images in the
coronagraphic image.
These calibrator images can be used for precise relative astrometry,
to establish companionship of other objects in the field of view 
through measurement of common proper motion or common parallax,
to determine orbits,
and to observe disk structure around the star
quantitatively. 
The calibrator spots also have known brightness, selectable by the coronagraph
designer, permitting accurate relative photometry in the coronagraphic image.
This technique, which enables precision exoplanetary science, is relevant to
future coronagraphic instruments, and is particularly useful for ``extreme''
adaptive optics and space-based coronagraphy.
\end{abstract}

\keywords{
 instrumentation: adaptive optics ---
 instrumentation: high angular resolution ---
 space vehicles: instruments ---
 techniques: high angular resolution ---  astrometry
 planetary systems
}


\section{Introduction}

Stellar coronagraphs suppress a star's point spread function (PSF) in order to enable the
imaging of faint objects or structure, such as planets or proto-planetary disks,
in orbit about the  star
\citep{Lyot39, Sivaramakrishnan01}.
Accurate relative astrometry between the star and its apparent companions
or extended disk structure is essential to establish physical association through
confirmation of common proper motion or common parallax
\citep{Oppenheimer018pcsurvey}, to determine orbital parameters,
and to observe disk structure quantitatively.  Relative photometry
between the star and these other objects in the field of view is also
necessary to conduct studies of disk or companion physics and chemistry.
However, a perfect coronagraph will completely remove the core of the star's PSF,
meaning that measuring its intensity or location in the scientific data is impossible.
\citet{LS05} show that decentering the on-axis star results
in complex PSF morphology --- tilt errors can even introduce artifacts
that look like point sources within a few resolution elements of the center
of the occulting focal plane mask of the coronagraph, even in simple theoretical
simulations.  We have also observed these effects in real data from
the optimized, diffraction-limited Lyot Project coronagraph
\citep{Sivaramakrishnan01,  OppenheimerAMOS03, OSM03,
OppenheimerSPIE04, Makidon05},
which is deployed at the Air Force AEOS 3.6\ m telescope \citep{LRCN02}.
This coronagraph is the first to operate
in the regime of  ``extreme adaptive optics'' (ExAO),
with Strehl ratios in the 70-85\% range,
and it has played a role in developing advanced techniques of exoplanetary
science.
(The NICMOS coronagraph on the Hubble Space Telescope is not optimized for
coronagraphy --- the function of its focal plane occulter is to reduce the total
flux on the detector, since there is no coronagraphic suppression of the stellar 
background away from the PSF core.)
It brings early results to a burgeoning field of
coronagraphic imaging of nearby stars.
In \citet{Digby06} we discuss some of the problems of coronagraphic astrometry
uncovered when analyzing real data from the Lyot Project coronagraph,
including  the fact that such data is often far more complex than simple
theory might predict \citep{LS05, SSSLOM05}.
Over exposures a few seconds to several minutes long  objects are smeared 
out on the detector because of slow drifts in the AO system, differential 
refraction effects and field rotation.  In addition, the dynamic range of 
the detector prevents simultaneous non-coronagraphic imaging of faint companions 
with the AO target star.  The image is further complicated by the motion of
speckles, which occurs on all timescales present in the obervation.
This paper presents a solution to some
of the problems discussed in \citet{Digby06}.

We explain how to create controlled fiducial spots
in the coronagraphic image which pinpoint the location of the star,
in addition to providing calibration for accurate relative photometry,
thus solving the fundamental problem of coronagraphic image calibration.
By inserting a reticulate grid of wires into a pupil ahead of the focal plane
occulting mask in the coronagraph
we create predictable images of
the central star at desired locations and chosen brightness
in the coronagraphic image plane.
We extend the analytical theory of segmented aperture telescope
point-spread functions (PSFs) for direct imaging
\citep{Chanan98, SivaramakrishnanGSMT01, Yaitskova02}
as well as that of coronagraphic imaging \citep{SivaramakrishnanGSMT01, SY05}
to describe the effects of our proposed reticulate grid of wires
in the coronagraphic case.

Our methods, though relevant to any coronagraphic observation, are
particularly applicable to space-based high dynamic
range coronagraphy dedicated to the extremely challenging task of finding
and characterizing Earth-like exoplanets, as our approach
does not disturb the extremely smooth, well-corrected wavefront that such
searches demand.   
Our technique does not require exquisite calibration of deformable mirrors,
for instance, nor does it require any change in a deformable mirror
shape before, during, or after an observation.  

We note a simultaneous but independent study by \citet{Marois06}, where
a pupil plane before the focal plane mask is also modified in order to create
images suitable for photometric and astrometric purposes.

\S 2 elucidates the theoretical basis for our calculations of telescope PSFs.
\S 3 describes the perfect coronagraph used to illustrate the utility of this new
theoretical construct, as well as the difficulties in proper numerical simulations
of the reticulate pupil mask.
In \S 4, we extend the theory to a more practical level, 
addressing the effects of finite bandwidth in astronomical imaging,
imprecise positioning of optics,
residual uncorrected wavefront errors, and atmospheric differential
refraction.
\S 5 describes the practical application of the theory, including 
some details on how to design
reticulate pupil masks, and how to calibrate and reduce
coronagraphic data to derive precision relative astrometry and photometry.

\section{Point-spread function theory}

In this section we describe our monochromatic Fourier optics formalism.
We recollect that a plane monochromatic wave
traveling in the $z$ direction in a homogenous medium without
loss of energy can be characterized by a complex
amplitude $E$ representing the transverse (\eg\ electric) field strength of the wave.
The full spatio-temporal expression for the field strength is
$E e^{i(\kappa z - \omega t)}$,
where $\omega/\kappa  = c$, the speed of the wave.
We do not use the term {\it field} to denote image planes --- 
the traditional optics usage --- we always use the term to denote
electromagnetic fields or scalar simplifications of them.
The wavelength of the wave is
$\lambda = 2\pi/\kappa $.
The time-averaged intensity of a wave at a point is proportional to $EE^*$,
where the average is taken over one period,  $T = 2\pi/\omega$,
of the harmonic wave.
The phase of the complex number $E$ represents a phase difference from the
reference phase associated with the wave.   A real, positive $E$ corresponds to 
an electric field oscillating in phase with our reference wave.  A purely imaginary
positive value of $E$ indicates that the electric field lags by a quarter cycle from the 
reference traveling wave.
Transmission through passive, linear filters such as apertures, stops and
apodizers is represented by multiplying the field strength by the
transmission of these objects which modify the wave.  Again, such multiplicative
modification is accomplished using complex numbers to represent 
phase changes forced on the wave incident on such objects.

We assume that the Fraunhofer approximation describes our imaging system.
Thus image field strengths are given by the Fourier transform of aperture 
(or pupil --- we use the two terms interchangeably) illumination functions,
and vice versa.

\subsection{The reticulated aperture point-spread function}  \label{segpsf}

Gaps between segments are analogous to the wires in our grid, and our
wire grid spacing replaces the segmentation periodicity.  
We use the notation of segmented apertures in order to facilitate
comparison of this work with existing work on segmented aperture PSFs,
but refer to wire thickness and spacing rather than segment gap and width.
We re-derive the form of square-segmented aperture PSFs
in both direct and coronagraphic imaging analytically,
extending the approach of \citet{SivaramakrishnanGSMT01, SY05}.

The full aperture of the telescope is described by a 
function $A(\bx)$, which, for unapodized apertures, is unity over
the aperture and zero elsewhere.  The position vector $\bx=(x,y)$ 
denotes location in the aperture plane, in units of the wavelength
of the light.
We develop the expression for the monochromatic PSF
with a perfect wavefront first, and then calculate the coronagraphic PSF,
assuming that the scalar wave approximation and Fourier optics provide
an adequate description of PSF formation.  We treat aberrated wavefronts 
in Section \ref{wfesensitivity}.

The aperture illumination function --- the complex amplitude in the aperture
in response to a unit (field rather than power) strength incoming wave ---
for the reticulated aperture, can be written as
	\begin{eqnarray}  \label{Aseg}
	A_s(\bx) &&\eq  A(\bx) ~ \times \nonumber \\
	  && \big ( 1 -  \shah((x-x_o)/d)~/~d ~ \conv ~ \Pi(x/\g) \big ) ~ \times \nonumber \\
	  && \big ( 1 -  \shah((y-y_o)/d)~/~d ~ \conv ~ \Pi(y/\g) \big ).
	\end{eqnarray}
Here $d$ is the wire grid's  spatial periodicity, $\g$ the 
wire width;
$\conv$ denotes the convolution operation.
We make use of Fourier analytic notation, techniques and results
which can be found in \eg\ \citet{Bracewell}.
The one-dimensional shah or comb function $\shah$ is defined by 
	\begin{equation} \label{shah}
	\shah(u) \equiv \sum_{n = -\infty}^{\infty} \delta(u - n),
\end{equation}
where $\delta(u)$ is the Dirac $\delta$-function,
and
    \begin{equation} \label{tophatdef}
	\begin{array}{ll} \Pi(x) \eq 1 & {\rm for\ } |x| < 1/2, \\
	                  \Pi(x) \eq 0   & {\rm elsewhere}.
	\end{array}
    \end{equation}
Normalization by $1 / d$ is required to ensure 
the correct impulse strength
in each $\delta$-function in equation~(\ref{Aseg}).
We offset the origin of the grid from that of the 
full aperture by an arbitrary vector shift of $(x_o, y_o)$,
restricting $x_o$ and $y_o$ to be less than the grid size $d$ 
(without loss of generality).
This lateral offset of the reticulate pattern is a detail 
that segmented aperture telescope PSF calculations
(\eg\ \citet{Chanan98, SivaramakrishnanGSMT01, Yaitskova03, SY05})
do not need to consider, but is relevant to our case, as it enables
analytical tolerancing of the in-plane mechanical stability and
placement of the wire grid.

The complex amplitude of the image resulting from the aperture 
illumination function $A_s$ is its Fourier transform,
	\begin{eqnarray}  \label{aseg}
a_s(\bk) && \eq  a(\bk) ~ \conv  \nonumber \\
&& \big ( \delta(k_x) -  \g ~\shah(dk_x)~e^{ik_xx_o} ~ \sinc(\g k_x) / d \big ) ~ \conv \nonumber \\
&& \big ( \delta(k_y) -  \g ~\shah(dk_y)~e^{ik_yy_o} ~ \sinc(\g k_y) / d \big ),
	\end{eqnarray}
where $\bk=(k_1,k_2)$ is the image plane coordinate in radians.
We adopt the convention of changing the case of a function to indicate its Fourier
transform.
We refer to this complex-valued field $a$ as the
``amplitude-spread function'' (ASF),
by analogy with the PSF of an optical system.
The ASF is the electric field in the image plane.
The PSF is the square of the absolute value of the ASF (equation (\ref{pseg}), below).
Using the definitions
	\begin{eqnarray}  \label{Gdef}
   G_x(\bk) & ~\equiv~ & - e^{ik_xx_o} \sum_{n = -\infty}^{\infty} a(k_x - n/d, k_y) ~ \sinc(\g k_x) \nonumber \\
   G_y(\bk) & ~\equiv~ & - e^{ik_yy_o} \sum_{m = -\infty}^{\infty} a(k_x, k_y - m/d) ~ \sinc(\g k_y) \nonumber \\
 G_{xy}(\bk)& ~\equiv~ &  e^{i(k_xx_o + k_yy_o)}  \sum_{n,m = -\infty}^{\infty} a(k_x - n/d, k_y - m/d) \nonumber \\
   && \times  \sinc(\g k_x) \sinc(\g k_y),
	\end{eqnarray}
		%
the expression for the ASF of the reticulated pupil is
		%
	\begin{eqnarray}  \label{asegIII}
  a_s(\bk) & \eq & a(\bk)  +  \e \big (G_x (\bk) + G_y (\bk) \big )  +   \e^2~G_{xy}(\bk) \nonumber \\
		%
	\end{eqnarray}
We use the dimensionless number $\e \equiv \g/d$ to characterize the 
wire width relative to the wire spacing.
The corresponding PSF is
	\begin{eqnarray}  \label{pseg}
 p_s(\bk) & \eq & |a|^2 +  \e^2 \big ( |G_x|^2 + |G_y|^2 \big ) +   \e^4~|G_{xy}|^2  \nonumber \\
 & + & \e~\big ( a(G_x^* + G_y^*)  +  a^*(G_x + G_y) \big ) + {\mathcal O}(\e^3),
	\end{eqnarray}
		%
We explicitly write the $\e^4|G_{xy}|^2$ term in
spite of the presence of a `catch-all' $3^{\rm rd}$ order
remainder term ${\mathcal O}(\e^3)$ because 
the former has peaks that could on occasion be significant in the
coronagraphic PSF.
At the origin the expression for $a_s$ in equation (\ref{asegIII})
reduces to $(1~-~\e)^2~a(0,0)$, so the peak value
of the reticulated aperture PSF is  $(1-\e)^4~p(\bk)$, or $ \sim(1~-~4\e)~p(\bk)$ if
$\e \ll 1$  (by the binomial approximation)
where $p = aa^*$  is the full aperture PSF.
The accuracy of this approximation affects our ability to estimate photometric
calibration analytically, but in any case the satellite PSFs will require 
their own calibration methods.
In practical applications $\e$ will likely be far smaller
than the value we use in our numerical example below
(see \S \ref{design}), so use of the binomial approximation
produces useful analytical insight by enabling a simple
theoretically predicted expression for the value of the PSF maximum.

\subsection {Components of the point-spread function}

\begin{figure*}[htbp]
\epsscale{0.5}
\plotone{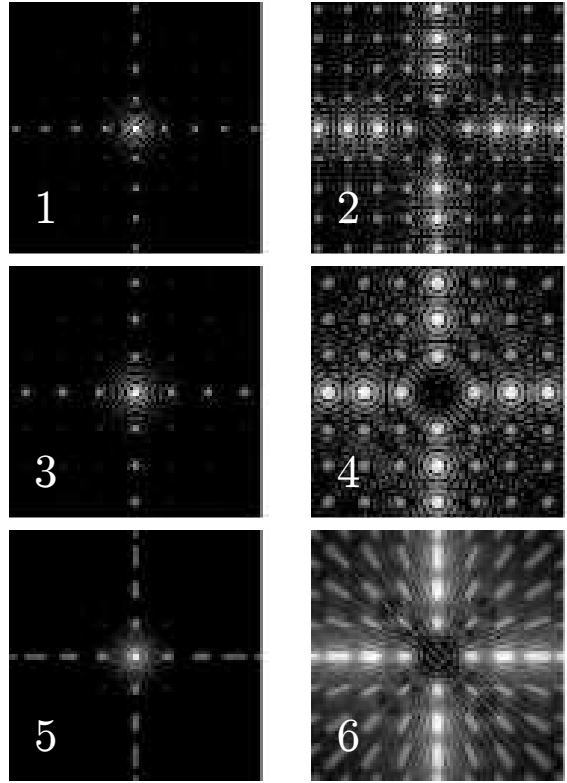}
\caption{ 
Monochromatic and broad-band direct and coronagraphic PSFs with a square-geometry
reticulate pupil mask.  All images are on a logarithmic greyscale 
stretching 10 magnitudes fainter than their peaks.  The pupil is 128 pixels across, 
and the grid has a wire spacing of 16 pixels, with 2-pixel wide wires.
	1: direct PSF for the shortest wavelength of a 20\% bandwidth filter
	with uniform transmission within the bandpass, in the absence in
	of phase errors.
	The satellite PSFs off the origin but along the horizontal and vertical axes 
	are fainter than the central core of the PSF by a factor
	$\e^2 = (\g/d)^2$, where $\g$ is the wire thickness
	and $d$ the wire spacing.
	The satellite spots off the axes are $\e^4$ fainter
	than the corresponding central peak.
	2: coronagraphic PSF at the shortest wavelength of the filter.  The off-axis
	sea of satellite spots are more visible in the coronagraphic image because 
	the core has been suppressed.
3 and 4: direct and coronagraphic PSFs for the longest wavelength of the filter.
5 and 6: direct and coronagraphic PSF for the full bandpass.
The length of any particular radial streak in this last pair of images,
(in resolution elements at the central wavelength of the bandpass)
is approximately the fractional filter bandwidth multiplied by the radial
distance of the spot at band center.
The streaks all point towards the origin, so the smearing has no
effect on astrometric precision according to Fraunhofer regime 
image formation theory.
We suggest using the four satellite peaks closest to the core
as fiducials for the position of the central occulted star
in coronagraphic images.
} 
\label{fig:coromw}
\end{figure*}

\begin{figure*}[htbp]
\epsscale{0.5}
\plotone{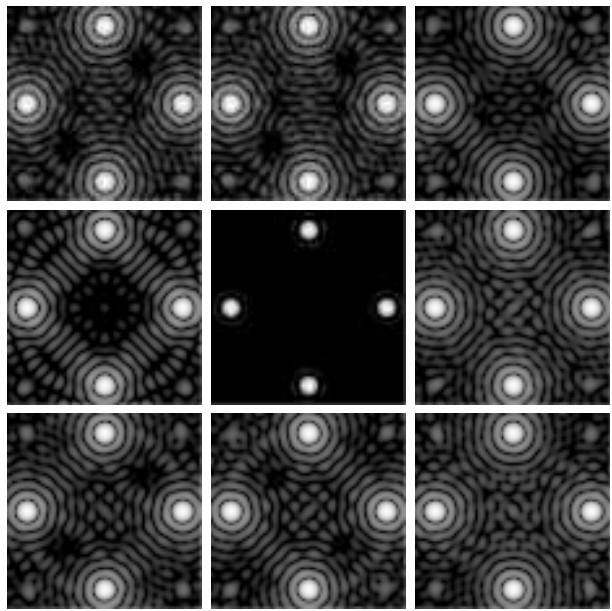}
\caption{ 
Monochromatic satellite spots in the coronagraphic image of a 
band-limited Lyot coronagraph.  
The origin of an exactly periodic wire grid in the pupil was
placed at 9 locations in a unit cell, chosen at random.
Completely symmetric satellite spots result, regardless of 
the in-plane placement of the grid. 
The logarithmic scale is 5 astronomical magnitudes for the central panel, and
10 magnitudes for the surrounding panels.
The grid was modelled with 16 wires, each two pixels wide, placed
across a binary circular unobstructed 512-pixel diameter pupil.
This wire is much thicker than that needed for a coronagraph
producing  a dynamic range of the order of $10^{6}$ or more.
}
\label{fig:corospots}
\end{figure*}

The PSF in equation (\ref{pseg}) can be decomposed into a few physically
significant components.
The PSF has a bright central core of $(1~-~4\e)~p(\bk)$. 
By inspection of the $\e^2|G_x|^2$ and $\e^2|G_y|^2$ terms
we see that the PSF possesses `satellite' copies of $p$  arranged 
along the arms of a cross, on two linear arrays
at a spacing $\lambda/d$ on the $k_x$ and $k_y$ axes,
with an overall taper by $\sinc(gk_x)$ and $\sinc(gk_y)$
respectively.
(Fig.~\ref{fig:coromw}, panels 1 and 3 are monochromatic, panel 5 shows
a 20\% bandpass polychromatic PSF).
The $\e^4~|G_{xy}|^2$ term in equation (\ref{pseg})
yields a sea of copies of $\e^4~p$
in a two-dimensional grid off the axes
in $\bk$-space, also the same $\lambda/d$ spacing.
This ${\mathcal O} (\e^4)$ component is not visible in 
the direct images in Fig.~\ref{fig:coromw} (panels 1, 3, and 5) because they are
too faint to be seen with the grey scale used in these panels. 
However, they are apparent in the coronagraphic PSFs using the same
reticulated aperture (Fig.~\ref{fig:coromw}, panels 2, 4, and 6),
where the central core has been suppressed.
There are first and higher order ``fringing effects'' in the PSF
(Fig.~\ref{fig:corospots}).
These are  described by cross terms in equation (\ref{pseg}).
The first order fringing term is $ \e(a(G_x^* + G_y^*)  +  a^*(G_x + G_y))$.
When $x_o$ or $y_o$ are non-zero these fringing effects can introduce
a complicated (but still centro-symmetric, since $A_s(\bx)$ is a real function)
morphology in the image.
The phenomenon of speckle pinning 
\citep{Bloemhof01, Sivaramakrishnan02, Perrin03, AS04ApJL}
will cause complex interactions with residual phase
aberrations in the brighter parts of the image,
so where these fringes overlap with the bright satellite
peaks their effects on astrometry and photometry will be more severe
than in the darker areas inbetween satellite peaks.
We discuss this further in Section \S \ref{wfesensitivity}.

If there are many grid spacings across the full aperture,
$\lambda/D \ll \lambda/d$, and the fringing between
various components
of the ASF is greatly reduced in the PSF.
In this case the PSF in equation~(\ref{aseg}) separates out
into obvious physical components deriving from the full aperture
ASF and the squares of the absolute values of $G_x$, $G_y$ and $G_{xy}$.
%
This entire pattern of full aperture PSFs is weighted by a
broad 2-dimensional sinc function envelope with a characteristic scale $\lambda/g$,
which is the widest angular scale in the ASF.  This scale derives from
the inverse of the smallest physical distance in the reticulate aperture,
\viz\ the wire thickness.
This overall weighting is almost constant at angular distances of $\sim 0.1~\lambda/g$
from the center of the PSF, simply because of the properties of the
$\sinc$ function.
Thus both sets of satellite PSFs (on the $k_x, k_y$ axes as well as off them)
have more or less uniform intensities near the core. 
This morphology is visible in Fig.~\ref{fig:coromw}.
As $\e \rightarrow 0$, $p_s(\bk)$ reduces to a single 
central PSF $p(\bk)$ at the origin ---
satellite spots and fringing disappear completely, as expected, when the wires vanish.

Alternate geometries can be created with different spacings in the
$x$ and $y$ axes when using specialized entrance apertures,
such as that currently envisaged for the Terrestrial Planet Finder.  In point of
fact any periodic two-dimensional wire grid can be used.

Before we discuss the utility of forming PSFs of this nature, we must derive
the coronagraphic PSF.  This is done in the next section.
\S \ref{bbimaging}, then, deals with the effects of finite
spectral bandwidth,
an imprecisely placed reticulate pupil mask and residual phase
aberrations in the entrance pupil.


%

\section{Coronagraphic PSF on a reticulated aperture}

\begin{figure*}[htbp]
\epsscale{0.75}
\plotone{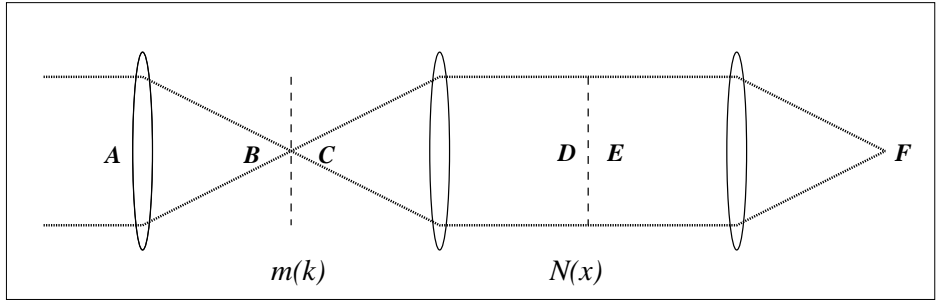}
\caption{
The essential planes and stops in a coronagraph.  The entrance aperture
is A, the direct image at B falls on a focal plane mask whose transmission function
is $m(k)$. The re-imaged pupil plane D, after being modified by passage through
a Lyot stop with a transmission function $N(x)$, is sent to the coronagraphic
image at F.  We place a square grid of opaque wires over the pupil plane A to create
controlled fiducial spots in the coronagraphic image at F for astrometric
and photometric purposes.
} 
\label{fig:corolayout}
\end{figure*}

A telescope aperture is described by a transmission function
pattern $A(\bx)$, where $\bx=(x_1,x_2)$ is the location in
the aperture, in units of the wavelength of the light
(see Fig.~\ref{fig:corolayout}).
The corresponding aperture illumination describing the electric
field strength in the pupil (in response to an unaberrated,
unit field strength, monochromatic incident wave)  is
		$ E_A =  A(\bx) $.
The field strength in the image plane, 
						$E_B = a(\bk)$,
is the Fourier transform of $E_A$
where once more,  $\bk=(k_1,k_2)$ is the image plane coordinate in radians.
Because of the Fourier relationship between pupil and image fields,
$\bk$ is also a spatial frequency vector for a given wavelength of light.
The PSF is $p = aa^*$,
recalling our convention of changing the case of a function to indicate its Fourier
transform. We, again, refer to $a$ as the ASF.
\placefigure{fig:corolayout}
We multiply  the image field $E_B$ by a mask function $m(k)$ to
model the focal plane mask of a coronagraph.
The image field immediately after this mask is $ E_C = m(\bk) \,  E_B $.
The electric field in the re-imaged pupil after the focal plane mask 
(the Lyot pupil) 
is $E_D$, which is the Fourier transform of $E_C$. 
We use the fact that the transform of the image plane field $E_B$
is just the aperture illumination function $E_A$ itself,
so the Lyot pupil field is
	$ E_D =  M(\bx) \conv  E_A $, 
where  $\conv$ is the convolution operator.

If the Lyot pupil stop transmission is $N(\bx)$, the electric
field after the Lyot stop is $E_E = N(\bx) E_D$.  The transform of this
last expression is the final coronagraphic image field strength:
		$ E_F  = n(\bk) \conv [m(\bk) \, E_B]$.
\citet{SY05} describe the structure of the field strength $E_D$
in the Lyot plane located at D.  We use their results
to analyze the final image plane $ E_F $ for an ideal (``perfect'') coronagraph
on a reticulated circular aperture, in order to use its features as 
fiducial locations in the final image for the purpose of high precision 
astrometry and photometry.

\subsection{The ideal coronagraph}

The band-limited coronagraph (BLC) design is mathematically ``perfect''
in that simple Fourier optics modelling predicts that it will
prevent all incoming, on-axis light from reaching the final coronagraphic
focal plane \citep{KT02}.
Its mathematical simplicity makes it useful for the purpose of
elucidating the way Lyot-style coronagraphic PSFs
are affected by various factors, such as
tip-tilt \citep{LS05},
higher order phase aberrations \citep{SSSLOM05},
the presence of secondary mirror support vanes or ``spiders''
in the pupil \citep{SL05},
the secondary obstruction itself \citep{Sivaramakrishnan04xaopi},
or inter-segment gaps \citep{SY05}.  The BLC design also
enables an analytical treatment of various effects, which can then be studied
numerically for particular instruments, or on related designs such
as the Apodized Pupil Lyot Coronagraph \citep{SAF03, Soummer05, SL05}.

The purpose of this exercise is that although the BLC may not
be used in any particular coronagraphic instrument,
the theory allows us to examine the behavior of the
reticulated mask independent of the particular choice of coronagraphic starlight
suppression.  In this way, our analysis is not complicated by the difficulties
of starlight suppression, by the simple fact that we assume it is perfect
in this analysis. 

The BLC design is easily understood if we introduce a ``mask shape function''
$w$ with the definition
	\begin{equation} \label{maskfunction}
		m(\bk) \equiv  1 - w(\bk),
	\end{equation}
which means that $M(\bx) = \delta(\bx) - W(\bx)$.  We use this
in the expression for the Lyot plane field strength $E_D$,
noting that since the focal plane mask is opaque at its center,
$w$ is unity at the origin, so $W$ is a function whose
two-dimensional integral over the whole of the pupil
plane is unity.
Thus we can write the electric field at the Lyot plane as
	\begin{equation} \label{lyotstopfield}
		\begin{array}{ll}
		E_D(\bx) & \eq  A \conv (\delta(\bx) - W(\bx)) \\
		              & \eq  A - A \conv W(\bx).
	\end{array}
	\end{equation}
If $w$ is a band-limited function with bandpass $b$,
then there is a minimum positive value of $b$ such that
the mask function's Fourier transform, $W$, satisfies the property
	\begin{equation} \label{bandlimit}
		W(\bx) \eq 0  {\rm \ if\ } |\bx| > b.
	\end{equation}
Thus in the BLC design $W$ (the Fourier transform of the mask shape function)
has compact support in its domain (the Lyot pupil plane).

In the case of a completely unaberrated coronagraphic optical train,
the above-mentioned properties of $w$ (or~$W$) result in the Lyot 
pupil field at any point further than $2b$ from a pupil edge being
identically zero.  We must also invoke the fact that the area under the
convolution of two functions is the product of the areas under the two
constitutent functions to see that this is indeed so.

In our numerical examples of coronagraphic images,
which are all unobscured circular aperture systems,
we use a BLC with a mask profile
	\begin{equation} \label{bandlimitNumeric}
		m(\bk) \eq   1 -  \jinc^2(|\bk|/k_o),
	\end{equation}
where $\jinc(x) = 2J_1(x)/x$.
Here $J_1(x)$ is the Bessel function of the first kind, with index $1$. 
The constant $k_o$ is chosen so that the first zero of the $\jinc$
function lies 12 resolution elements ($\lambda/D$)
from the center of the opaque focal plane mask.  Since this is about
10 times as wide as the direct image's PSF,
the bandpass of this $\jinc^2$ mask is approximately $D/5$
(the bandpass of a
product of two functions being the sum of the bandpasses of the
two functions, 
the bandpass of our jinc function is $D/10$).
Therefore the optimally-sized Lyot stop for this ``fourth-order''
focal plane mask is $3D/5$ in diameter  (the order here refers to the
local angular intensity profile of the suppressed or nulled area of
the coronagraph).

\subsection{Numerical simulations}

Our numerical example in Fig.~\ref{fig:coromw} uses a 128-sample wide 
binary pupil in an array of 1024 elements on a side, wires 2 samples wide,
with 8 wire gaps across the pupil in the $x_1$ and $x_2$ directions.
Those in Figs.~\ref{fig:corospots}, and \ref{fig:corowfe} use a 512-sample
pupil embedded in a $2048 \times 2048$ array, wires also 2 samples wide,
and 32 samples between wires, resulting in 16 wire spacings across the pupil.
Our fast Fourier transforms in double-precision (64-bit) numbers 
are implemented in the FFTW package \citep{fftw} wrapped in Python's
numarray package \citep{numarray02, pycon03}.

The lattice of satellite spot locations
in our image plane are at $(16 n \lambda/D, 16 m \lambda/D)$,
where $n$ and $m$ are integers (at least one of them being non-zero).
Two pupil samples across each wire
are barely adequate to sample the wires, although this suffices
for illustrative purposes.
However, there is a clear separation
of scales between our wire thickness and wire spacing with 
these parameters.  The wire thickness to wire spacing ratio, $\e$,
is 1/16, so our first few satellite spots along the $k_x$ and $k_y$
axes are $\e^2 = 3.91 \times 10^{-3}$ (6 magnitudes) fainter
than the central PSF core in the direct image. 
Our coarsely-sampled simulations confirm the theoretically calculated
intensity ratios to within 4\%.

However, the spots in the coronagraphic image are 
a factor of $5/3$ wider (in linear extent) than the direct image's
spots, because the optimized Lyot stop is $3/5$ as large
as the entrance pupil.
This also means that the spots, while the same angular distance from
the core, are $3/5$ closer in terms of the effective resolution
of the system because of the Lyot stop undersizing in this 
Lyot coronagraph design.
This change in size is dependent on the choice of starlight suppression, 
and is not a property of the reticulate pupil mask.  
Computers available now are prohibitively slow at simulating a two-dimensional grid of 
fine wires across a pupil at the scales of interest to the exoplanet hunter,
so analytical estimates are useful for understanding the behavior of this configuration.  
For example, to create fiducial calibration spots on the order of the brightness of
a planet, so that the calibration does not overwhelm the planet's brightness,
one would want fiducial spots of the order of 15 magnitudes fainter than the star,
which is approximately the $H$-band flux ratio between a 40 Myr old Jupiter-mass
companion and its host star \citep{Burrows04}.
Thus, the wires would need to be one thousandth as thick as their spacing.
Placing two numerical samples across the wires in this case would
entail $32768 \times 32768$ pupil arrays.
Typically coronagraphic computations require oversizing these arrays
by another linear factor of 4 to 8 in order to provide sufficiently fine
sampling in the pupil plane.  Thus numerical FFT methods will need to
handle arrays 196,000 samples on a side.  While direct PSF calculations
involving the semi-analytical `grey pixel approximation' do not require
such array oversizing \citep{Troy03}, this approximation may not be accurate
for coronagraphic modelling.

The results of our crude calculations of reticulate mask behavior 
are shown in Fig \ref{fig:coromw}.  The left hand column (panels 1 and 3)
are direct images of a star with the reticulate mask in place.
The right column (panels 2 and 4) shows coronagraphic images.
Panels 1 and 2 are at the shortest wavelength of a typical astronomical
filter's bandwidth ($\sim 20$\%), while panels 3 and 4 show images made
at the longest wavelength.  Panels 5 and 6 are broad-band images.
We now describe the derivation of these broad band images, as well as the
technique's sensitivity to optic placement and residual wavefront error.

%

\section {Broad-band theory and tolerancing}  \label{bbimaging}

The theory and numerical models we present in the preceding section
apply to the case of monochromatic light.  A comparison of the two upper sets of
panels in Fig.~\ref{fig:coromw} shows that the fiducial spots to be used for 
astrometric and photometric calibration are at (angular) positions in the focal plane
that are a function of wavelength.
Indeed, the spectral width of an observing bandpass induces radial
smearing of the fiducial spots. 
These spots will be miniature, low-resolution spectra of the central, obscured star.  
Intuitively, this is due to the fact that the reticulate mask acts as an extremely
coarse grating inserted into the beam.

For the purposes of astrometry and photometry,
we are interested in the four closest satellite PSFs located on the
$k_x$ and $k_y$ axes.
These four spots lie at $(\pm\lambda/d, 0)$ and  $(0, \pm\lambda/d)$
relative to the central star.
At this point we ignore any higher-order spots, assuming that the coronagraph
that uses this technique will be designed so that this first set of spots are
placed as far away from the central star as possible, while still remaining
in the field of view, so as to obscure as little of the potential search area
around the star as possible.  In \S \ref{lateralslip} we argue that there
are benefits that accrue if the satellite spots lie within the AO control
area of the wavefront correction system used in tandem with the coronagraph.

\subsection {Satellite spot intensity}  \label{bbexposure}

The four spots closest to the star from a reticulate pupil with $N = D/d$ 
wire spacings across the full pupil (of diameter $D$) will lie at 
	$(\pm N \lambda/D,  0)$ and 
	$(0,  \pm N \lambda/D)$ 
in the image plane at any particular wavelength $\lambda$.
If the observing bandpass, with central wavelength $\lambda_c$, has
a fractional bandwidth
$ \beta \equiv \delta \lambda / \lambda_c$,
the radial smearing of one of these spots is 
	$N \delta \lambda / D = N (\delta \lambda / \lambda_c)\lambda_c/D$,
	or about $N \beta \lambda_c/D$.  The spots themselves extend over the coordinates
	$(\pm N \beta \lambda_c /2D,  0)$ and 
	$(0,  \pm N \beta \lambda_c /2D)$.
Approximating the equivalent width of PSFs in the entire
spectral band by that of the PSF at the central wavelength
of the bandpass, $\lambda_c$, we see that each radial smear
covers approximately $1 + N \beta$ square resolution elements.  
Fig. \ref{fig:coromw} (panels 5 and 6) shows a numerical simulation of 
a broad-band PSF with a 20\% bandwidth.
We assume a perfect detector and a perfectly flat stellar spectrum,
with sharp cutoffs at both the long and short wavelength ends
of the filter pass band in these figures.

These spot widths will depend on the type of coronagraph.
In classical unapodized coronagraphs (\eg\ \citet{Sivaramakrishnan01, SY05})
the satellite image width is determined by the effective pupil diameter of the
system {\it after} the undersized Lyot stop. 
This affects the final image size by effectively reducing $D$,
though the reticulate mask will act on the entrance pupil of size $D$.
Lyot pupil undersizing affects the width of the satellite spots 
rather than their angular location in the focal plane.
This detail must be taken into account when estimating the satellite
PSF brightness, for example, in order to design a reticulate mask
for a given application.

The total flux in the four closest satellite
images is $4 \e^2 $ of the stellar on-axis flux in
the absence of a focal plane mask, but through the undersized
Lyot stop.  Using this for accurate photometry is discussed in \ref{photom}.
We show in \S \ref{astrom} that the fact that each smeared spot is 
a stellar spectrum does not affect the precision of the astrometry.

\subsection {Sensitivity to in-plane positioning}  \label{lateralslip}
\placefigure{fig:corospots}

In order for this reticulate mask to be useful for astrometric and 
photometric measurements, we must show that slight misalignments of
the reticulate mask will not affect whether the star can be accurately located,
nor will they grossly affect the fiducial spot intensities.
Here we show that the characteristics of the spots relative
to the star are in fact determined by the reticulate mask
characteristics, not its position in the pupil plane.
We quantify the limits of this statement below.

\subsubsection{Photometric effects}

The number of wire grid spacings across the aperture, $N$,
determines the relative magnitude of the cross terms caused by
interference between  satellite peaks.
This parameter feeds into the PSF structure through 
$d = D/N$, thus affecting the separation between satellite spots.
We have already referred to this interference between satellite peaks
as fringing. 

A coronagraph suppresses the central peak of the PSF, but lets
the satellite peaks through to the coronagraphic focal plane
(see \citep{SY05} for details).
Therefore in the coronagraphic case fringing only occurs between
satellite peaks.  Fringing between nearest neighbor satellite peaks
on the $k_x$ or $k_y$ axes yield field strength amplitudes within the range
$\e a(0,0) \pm \e a(0,0)  N^{-\frac{3}{2}}$.
We have used the fact that at a distance of several
resolution elements from the origin, $a(\bk) \sim k^{-\frac{3}{2}}$
for a circular entrance aperture telescope by virtue of the asymptotic
behavior of Bessel functions (\eg\ \citet{AbramowitzStegun}).

Thus the absolute PSF intensity variation of our satellite peak due to 
fringing with its nearest neighbor is of the order of 
	$\e^2 a(0,0) (1 \pm  2N^{-\frac{3}{2}})$.
Remembering that the coronagraphic satellite peaks on the axes
have an intensity $\e^2 p(\bk)$, the fractional change in satellite PSF
brightness is therefore of the order of 
	$2N^{-\frac{3}{2}}$.

If $N \sim 25$, the resultant fringing is of the order of $1/64$, or 1.6\%,
of the satellite image's peak intensity.  For $N = 50$ this reduces 
to fluctuations of at most 0.6\%.
Examples of the fringe morphology are shown in Fig \ref{fig:corospots}.

As a result, arbitrary decentering of the reticulate mask affects photometric calibration
by no more than 1.6\%  in our example.  Although this places a limit on the
photometric precision attainable with this method, far superior performance can be achieved
by requiring a modest level of alignment between the reticulate mask and the pupil.
In any case, 1.6\% photometry of any exoplanet is certainly considerably more accurate than
any published exoplanet photometry to date.

When phase aberrations are present the ASF is given by 
$a_s(\bk) * FT(e^{i\phi})$. 
If satellite images are separated by more than the size of a seeing disk,
their peaks will track the Strehl ratio of the core closely.
This can be seen by the fact that $a_s(\bk)$, the perfect ASF,
is basically a set of $\delta$-functions on a grid with spacing $\lambda/d$.
This ASF is convolved by the Fourier transform of $e^{i\phi}$, the latter 
being the seeing disk.
Given that the value of $FT(e^{i\phi})$ at the origin tracks
the Strehl ratio exactly, each satellite peak will do so also,
as long as  $FT(e^{i\phi})$ drops to a negligible value at a distance
$\lambda/d$ from its origin, \viz\ the seeing disk is significantly smaller than
the satellite spacing.  When satellite images lie in the wings of the PSF
the situation is more complicated because the closest satellite peaks will 
lie on a pedestal of a seeing halo around the core. 
Of course the addition of the halo and the wings is coherent
rather than incoherent.
This situation needs further analysis to understand the photometric
correction of satellite intensities due to the core PSF's halo.
Given that an ExAO system reduces the parent full aperture PSF's
intensity to some noise floor within the AO control area, and the
satellite peaks must rise above this noise floor in order to be
identified and measured, the AO-corrected core will
not affect the satellite PSF photometry by more than the ratio of the satellite
intensity to the noise floor induced by residual phase errors.  The effects
of the nearest neighbor satellite peak will be down a factor of $\e^2$ from
the effect of the core on the satellite.  This will have a negligible effect
on the photometry of the satellite images.

\subsubsection{Astrometric effects}

From a simple theoretical standpoint, the locations of the spots will
always remain symmetric about the occulted star's position on the detector
in the absence of phase errors and differential refraction
through the atmosphere or transmissive optical elements.

A coronagraphic imaging survey designed to detect exojupiters over
a one to four year period requires astrometric precision on the order
of a milliarcsecond, or $\lambda/50D$ on an 8~m telescope in the $H$-band.   
Terrestrial Planet Finder science goals and unpublished studies  
conducted by both the Gemini and VLT planet imaging teams require 1  
to 10 mas precision in relative astrometry for discrimination between  
background objects and real companions of nearby stars when orbital  
motion is included.
There is no reason why  $\sim$  1/100th of a PSF width would not be achievable,
particularly with four fiducial spots.
The astrometry of the putative planet relative to the  
central star will be predominantly determined by the
statistical significance of the planet detection.

A tilt of the mask to the beam results in a diminution of the effective
value of $d$, since the projected separation in the pupil plane will contain
a factor of the cosine of the tilt angle.  
A tilt of 12 degrees is required to produce a projected separation change
of a factor of $1 - 1/50$, so mask tilt in a collimated beam does not
need to be tightly controlled.
If the reticulate mask is allowed to rotate as well as translate,
our method is not in principle sensitive to the azimuthal orientation
of the grid either, as long as our data reduction algorithms allow for
such rotation of the pattern of fiducial spots.

Ground-based systems using the reticulated pupil mask for astrometry
will also have to contend with differential refraction from the atmosphere.
Even with an atmospheric differential refraction corrector (ADC) there will
be residual differential refraction.  This will distort the small spectral
smears of  the satellite spots.  We have not estimated the magnitude of this effect,
as it depends on particular details of ADC design, as well as atmospheric 
properties \citep{Roe2002}.  By observing a star at different zenith distances
the residual errors due to atmospheric differential refraction can
be calibrated out to some extent.  This calibration may also be dependent on 
the spectrum of the star.  Furthermore, too much reticulate mask rotation
is likely to complicate the calibration of atmospheric differential refraction.


We conducted numerical simulations of
in-plane placement errors.
The nine coronagraphic images in 
Fig.  \ref{fig:corospots}
show the effect of random decenterings of the reticulate mask
within the unit cell of the grid pattern (a square of side $d$).
We use random placements to avoid potential `resonances' between
predetermined offsets that are themselves in a periodic grid
within the unit cell of the wire grid.  Simulation of many more 
random placements, in addition to systematic incremental grid
decentering, indicated that the first nine cases, shown here,
produce generic behavior.
Given the centro-symmetry of the coronagraphic images 
(which ensues since the pupil illumination function is a real function),
no change in the estimation of the center of the pattern will result if 
the astrometric data reduction method utilizes this fact.
In the absence of aberrations and photon noise,
in-plane grid placement errors make no 
intrinsic contribution to astrometric
errors, unless image distortion calibration errors are significant enough to destroy the
symmetry inherent in using a reticulate wire grid in the pupil.
Thus image distortion needs to be characterized to at least the same level
as the required astrometric accuracy demanded by the science.

More sophisticated wave propagation and geometric distortion models are needed 
to quantify astrometric sensitivity to lateral placement errors of the
grid in any particular instrument.  In the distortion-free Fraunhofer regime
lateral placement of the reticulate grid has no effect on the estimate of
the occulted star's position in the coronagraphic image.


\subsection {Sensitivity to residual phase aberrations}  \label{wfesensitivity}
\begin{figure*}[htbp]
\epsscale{0.75}
\plotone{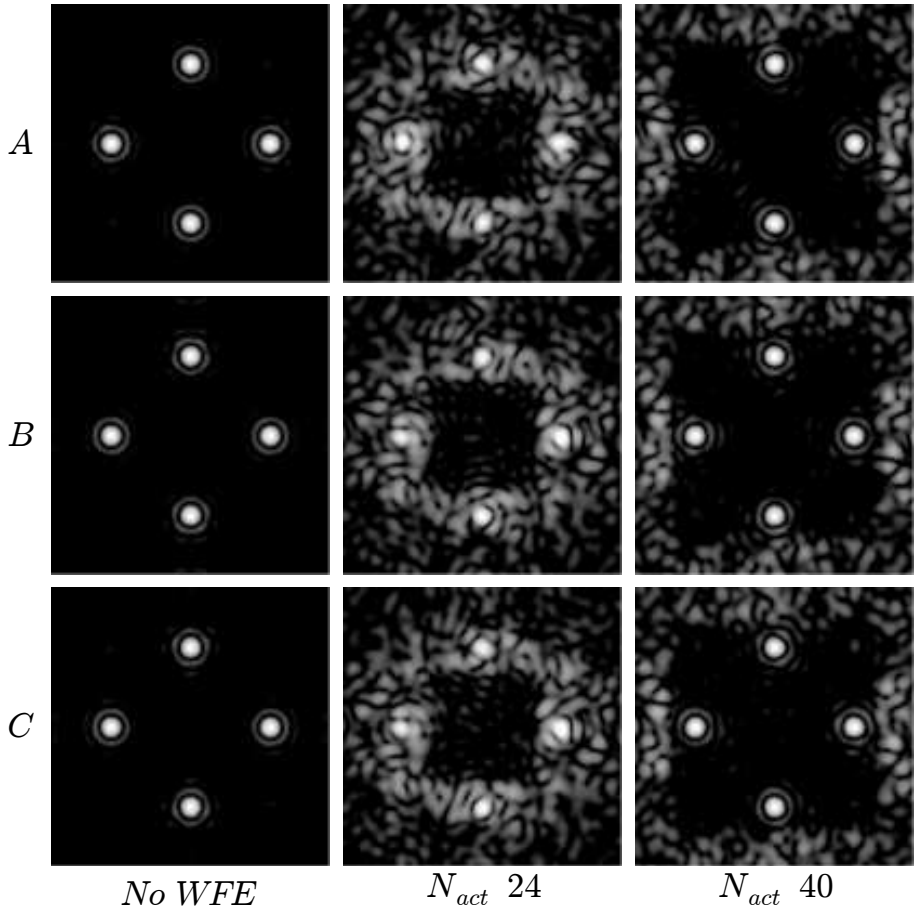}
\caption{ 
The effect of wavefront errors on astrometric fiducial spots.
Monochromatic satellite spots in the coronagraphic image of a 
band-limited Lyot coronagraph are calculated for wavefront errors 
that would produce a Strehl ratio of 95\%.
In each of the three rows (labelled A, B and C)
the origin of an exactly periodic wire grid in the pupil was
placed at 3 locations, chosen at random, in a unit cell. 
The ``No WFE'' column shows resulting coronagraphic
PSFs on a logarithmic stretch between the peak satellite PSF intensity
of $3 \times 10^{-4}$ (of the reference unocculted PSF using the 
the corongraphic Lyot stop), and a factor 251 times fainter
(6 magnitudes) below the satellite PSF intensity peak, and
14.8 magnitudes fainter than the central occulted star.
The simulated AO system (with spatially-filtered wavefront sensing)
used in the $N_{act}\  24$ column has
24 sensing and control channels across the entrance pupil diameter,
and the $N_{act}\  40$ column images result from AO correction with 
40 channels across the pupil, with the phase rescaled to possess the same variance
as the $N_{act}\  24$ case.
The reticulate grid in the entrance pupil was modelled in the same
manner as for Fig. \ref{fig:corospots}.
Residual speckles from uncorrected phase aberrations complicate the spots' 
morphology if the spots lie outside the dark square of the AO control region.
This suggests using the right-hand arrangement, where fiducial spots are
placed inside the AO control area for the most accurate astrometry.
%
%
}
\label{fig:corowfe}
\end{figure*}

\placefigure{fig:corowfe}

As \citet{Perrin03} point out, when pupil-plane phase aberrations
are the dominant aberrations, the PSF of an imaging system
can be decomposed into an infinite but absolutely convergent
series involving a combination of the perfect system's response and
the Fourier transform of residual phase aberrations  over
the pupil.
This work suggests that the fiducial spots' morphology
can be changed in complicated ways by these interactions, 
even in the case when the pupil-to-image relation is described by 
simple Fourier optics.
In light of these facts,
we examine how residual phase aberrations might affect 
the astrometric performance of the fiducial spots created by
a reticulate grid in the pupil.

The left column of Fig. \ref{fig:corowfe} (labelled ``No WFE'')
shows three realizations
of the unaberrated coronagraphic response of a perfect 
coronagraph, with the grid origin placed at three random locations
in the entrance pupil of the telescope.
This simulates in-plane placement errors of the mechanism holding the grid.

This coronagraphic configuration is illuminated with aberrated
but ExAO-corrected on-axis starlight.  
We simulate ExAO correction by high-pass filtering a realization of
a Kolmogorov spectrum phase screen, with $D/r_o = 15$.
The high-pass spatial frequency filter mimics
the action of a perfect ExAO system equipped with spatially-filtered
wavefront sensing (SFWFS) \citep{Poyneer04}. 
We generate a phase screen using the method described by \citet{Lane92},
and our numerical filtering procedure follows that of
\citet{Sivaramakrishnan01, Poyneer04, Makidon05}.
Our spatial filter blocks all spatial frequencies
below the filter's cutoff, but passes all signal at spatial
frequencies above the cutoff.  The cut-off spatial frequency
is determined by the number of actuators across the telescope
diameter (as long as wavefront sensing is performed
on at least as fine a spatial sampling as the control).
This frequency is referred to as the ExAO control frequency 
(\cite{Perrin03, Poyneer04} discuss this topic further).
The wires in our grid are spaced by $d = D/32$. 
 We then simulate two different ExAO systems,
one with 24 actuators across the entrance pupil ($N_{\rm act} = 24$), 
which places the fiducial spots well outside the ExAO control area,
and another in which $N_{\rm act} = 40$ to ensure the spots are located
inside the ExAO control area.
This allows us to evaluate how residual speckles
in the final image affect the utility of the satellite calibrator spots.
Coronagraphic PSFs from three realizations of grid placement in the pupil
of these two AO systems are shown in Fig \ref{fig:corowfe}.

After the spatial filtering simulating the ExAO, we rescaled the our phase 
samples to ensure that the variance of the filtered phase array is 
0.05 ${\rm radians}^2$, to ensure a Strehl ratio  similar to that expected 
for real ExAO systems being designed now.
The Mar\'echal approximation \citep{marechal} states that
this variance is the Strehl ratio reduction, so a 95\% Strehl ratio
image is placed on our coronagraph's focal plane mask.
Thus 5\% of the energy in the AO-corrected wave is distributed
in remnant speckles.
This energy leaks through the coronagraph
\citep{Sivaramakrishnan01,LS05,  SSSLOM05}, and affects the shape
and symmetry of the fiducial spots in the coronagraphic image plane.

The middle column of Fig. \ref{fig:corowfe} (labelled ``$N_{\rm act}$ 24'')
shows the same three realizations of pupil grid geometry as in the 
``No WFE'' column,
but with our Kolmogorov-spectrum phase screen corrected by an
ExAO system with 24 actuators across the entrance pupil.
With this arrangement the fiducial spots lie in the midst of
uncontrolled residual speckle.
The peak intensity of the four satellite spots in this image
is at a level of $\sim 3 \times 10^{-4}$ of that of the unocculted
star.
We use a logarithmic grey scale between this
peak and 1/251 of this peak (6 magnitudes of dynamic range)
in the figure.
The darkest regions of our images are 14.8 magnitudes 
fainter than an unocculted star.
At this display stretch we see that the fiducial spots
lose their simple morphology, as well as their symmetry,
when immersed in a sea of residual speckles.
%
%

The right column of Fig. \ref{fig:corowfe} (labelled ``$N_{\rm act}$ 40'')
shows the same three realizations of pupil grid geometry as in
the left column, but with 40 actuators across the entrance pupil.  
Now the fiducial spots lie inside the darker ExAO control
area, and are less prone to the deleterious effects of remnant
speckles than the $N_{\rm act} = 24$ case shown in the middle column.
We discuss these practical matters in the next section.

\section {Application of the Theory}  \label{cal}

The previous sections describe a new reticulate mask that produces fiducial
spots for astrometric and photometric calibration of coronagraphic images.
Using the Fraunhofer approximation of Fourier optics,
we have shown analytically, as well as with numerical simulations,
how these spots are formed in both monochromatic and broad-band light. 
We have also demonstrated that their brightness and spacing
is easily controlled, and that
extremely precise positioning of the reticulate mask is
not required for the use of the satellite image spots for
astrometry or photometry, and
furthermore, that our technique is usable with real ExAO corrected starlight.
In this section we explain how to design the mask, how it would be used,
and how astrometry and photometry can be derived from real data using our
reticulate pupil mask.

\subsection {Reticulate mask design}  \label{design}

As with many new concepts, there are several instrument-specific
details that must be addressed by an initial design.
Further refinements of any early design will doubtless occur
during the manufacture and use of a reticulate
pupil mask on a telescope.  Practical problems such as the quality of
pupil image at the grid will have to be studied --- this particular problem
appears to be tractable, since coronagraphs typically require good pupil
images, and beams in coronagraphic instruments have
high focal ratios, and, thus, slowly-converging beams 
\citep{OppenheimerAMOS03, OppenheimerSPIE04, bmac2004SPIE, mouillet2004ASPC}.

Residual speckle noise in a coronagraphic image sets the
contrast ratio floor that the instrument can achieve unless
post-processing methods (\eg\ \citet{Marois00, Marois04phd, Marois04})
can suppress this speckle noise further.  Here we assume coronagraphic
data is not processed further in such ways.
\citet{Hinkley06} report residual speckle noise 
in the $H$-band at the edge of
the AO control area 
($17 \lambda/D$ from the PSF core, or $1\arcsec.5$ at the $H$-band central wavelength)
in the Lyot Project coronagraph at $\sim  11$
magnitudes fainter than the central star.  We might therefore wish
to create fiducial spots that are clearly visible, but not
overpoweringly bright, near the edge of our usable AO-corrected
field of view.  The location of the fiducial spots is determined by the
periodicity of the grid of wires relative to the full 
pupil diameter, and the spots' brightness is set by the wire thickness
relative to the wire spacing, $\e = \g/d$.  With a 100\ mm pupil at the beam capture 
mirror BCM of the Lyot Project coronagraph, a wire spacing of 5.88~mm will place
the fiducial spots at $17 \lambda/D$ (1\farcs5 at $H$-band center)
along the grid's symmetry axes in the image plane.
In order to make the spots 10 magnitudes fainter
than the central star, we require a wire thickness of
$1/100$ of their spacing, \viz\ $58.8$  \micron\  diameter.

In summary, the relations describing the design are that the 
four closest and brightest satellites spots
are placed  $(D/d)(\lambda/D) = \lambda/d$ distant from the central star 
($d$ being the wire grid spacing as projected on to the entrance aperture),
and their intensity is a fraction  $(g/d)^2$ of the central star's intensity
($g$ is the wire thickness).  Note that $(D/d)$ is the number of wire spacings
into which the entrance pupil is sliced in each direction along the grid's axes.

\subsection{Observing strategy}

Our theoretical results suggests that fiducial spots should be
placed in areas of the highest contrast of the image, where 
there are fewest remnant speckles, if the most accurate
astrometry is desired.  However, we recognize that they would then
occupy a part of the optimal search region in the coronagraphic image.
An appropriate observing strategy can remove this problem, or the observer
can simply choose to ignore the region under the spots.  
If the spots are placed outside the AO control area and a significant amount
of speckle noise is expected around the spots, thicker wires should be
used in order to increase spot brightness. 
That way accurate astrometry could still be performed using the spots.  
The pupil mask can remain in place during observations,
and yet not obscure the darker regions of the focal plane.
Detectors that are destined for use in the next generation 
of ExAO systems typically utilize 2048-square arrays.
If these detectors are used in purely imaging modes
they can cover a wide enough field of view to enable the satellite spots
to fall outside the AO control area.
However, if an integral field spectrograph configuration
is used, the number of pixels required to cover the dark region
of the focal plane in both spectral as well as spatial
dimensions becomes a concern with these detectors.
In such cases the satellite spots can be placed within the
AO control area.
%
%

The astrometric device we describe could be left in the light
path of the instrument all the time, or it could be removed from the
beam, to be reinserted as needed.  The presence of these fiducial spots 
may hamper the study of faint diffuse structure around the target star,
and may also occupy too much of the best search areas in the coronagraphic
image. The reticulate mask must be positioned at an angle that is not coincident
with the telescope spider structure, or the calibrator spots may be lost in the
glare of the diffraction spikes,
and the spots' morphology confounded by residual spider diffraction \citep{SL05}.

The fiducials spots' locations and morphology will need calibration.  Binary 
and triple star systems, or fortuitous groups of stars with appropriate
brightnesses will need to be observed.  Given that the light diffracted
by the grid will diverge from the light of the parent star, starting at the 
location of the grid, a consideration of geometrical image distortion
is merited on a case-by-case basis for each instrument that uses this device.
There may be some utility to performing astrometric calibration of
the spots without the coronagraphic focal plane mask in place, but with
the undersized Lyot stop on classical coronagraphs.

\subsection{Relative astrometry}\label{astrom}

With four fiducial spots in the instrument's field of view, the spots 
provide most information about the star location if they are
treated together.  Due to the spectral smearing in the radial direction,
each set of two spots lie on a line intersecting the central star.
These two sets of two spots can be used independently to pinpoint the star
in each of the two orthogonal $k_x$ and $k_y$ directions, 
in the absence of atmospheric differential refraction. 
In actual data reduction, we suggest fitting a line to each set of two radially
smeared spots.
The fits of these spots should be achieved in a fashion similar to that used
for optimal extraction of spectral data from digital images.
(e.g., \cite{Opp98}, and references therein.)

The intersection of the two lines from the two sets of fits pinpoints the star.
This technique is independent of the spectrum of the star, since for each fit
only the shape of the smear PSF in the direction perpendicular to the spectral
smear is used.  As a scientific bonus, a very low-resolution spectrum
of the star can be recovered from each observation using the reticulate mask.
The ultimate precision of this technique is limited by photon noise.

The uncertainty in judging where the center of two radially-smeared
spots lie
(in the direction perpendicular to the spectral bandwidth-induced smearing)
is of the order of the size of the pattern in this direction ---
\viz\ a spatial resolution element --- divided by the square root of the number
of photons in the two satellite images (assuming the spatial resolution element
is at least sampled at the Nyquist frequency):
\begin{equation}
 	\sigma_x = \frac{\lambda_c}{D} \left( 2 \left( \frac{\g}{d}\right)^2 \int_{\lambda} \lambda Q_\lambda \frac{F^*_{\lambda} d\lambda}{ h c} \right)^{-1/2}  .
\end{equation}
Here the stellar spectrum is $F^*_\lambda$, and the detector
has a quantum efficiency as a function of  $\lambda$ of $Q_\lambda$.
$h$ is Planck's constant, 
and $c$ is the speed of light in a vacuum.  This assumes that the spots
are significantly brighter than the surrounding speckles.

\subsection{Relative photometry}\label{photom}

In conducting the optimal extraction fits of the smeared spots as described
in the previous section, a very coarse spectrum of the star is retrieved four times.
The sum of the counts in these four spectra is 
 $4\e^2$ multiplied by the stellar intensity (unocculted
by the coronagraphic mask, but including the throughput effects of the coronagraph).
The derived intensity can then be used in all images
with the reticulate mask in place to calibrate any other object in the field of view,
either relative to the primary star, or in an absolute sense if the primary star
is well calibrated and measured separately as a flux standard.

\section {Discussion} 

We have shown that a robust, easily-made optic can be incorporated 
into existing coronagraphs, or designed into future ones, in order to 
create controlled fiducial spots in the coronagraphic 
image plane for quantitative calibration.
This device is easy to use (in theory), and, furthermore, does
not require changing the phase of a wavefront with a
deformable mirror.  The latter quality could simplify the way
coronagraphic astrometry and photometry on a space-based optimized
coronagraph is performed.
At the limit, on the planned ExAO coronagraphs being constructed
over the next five years or so, we expect about 1/100 of a pixel
in astrometric accuracy for the primary, and  2\% photometry.
The accuracy of the relative astrometry between
the apparent companion and the coronagraphically-suppressed
central star depends entirely on the statistical significance of the
companion detection.
Fiducial spots placed inside the
region of the image plane controlled by the deformable mirror
are best for astrometry; the optic that generates the spots can be
removed and replaced as needed since astrometric sensitivity to
tilts and lateral shifts of the optic are low.
Laboratory and on-sky use of such a device will
develop our understanding of its behavior and utility.
\citet{Marois06} present early laboratory test results on the
behavior of a similar type of fiducial spot.
We plan to implement our reticulate pupil mask in the Lyot Project
coronagraph in the near future, for on-sky tests of the utility of this
technique.

If the wires in the grid are of the order of a few wavelengths of light wide,
these simulations may not apply,
and one must abandon the Fourier optics treatment and 
use the Fresnel approximation or full electromagnetic propagation methods
to calculate the expected location and 
characteristics of the fiducial spots.  Curvature and distortion
of optical planes --- for instance, the image of the pupil where we
place our grid --- might require our grid geometry to depart from its
simple periodicity in order to create compact, well-understood satellite spots.
For extremely demanding applications in future instruments these issues
will have to be addressed.

\acknowledgements
We acknowledge stimulating discussions with Christian Marois, Bruce Macintosh,
Ron Allen, and Stefano Casertano, and helpful comments by the reviewer.
The Lyot Project is based upon work supported by the National Science Foundation
under Grants AST-0334916, 0215793, and 0520822, as well as grant NNG05GJ86G
from the National Aeronautics and Space Administration under the Terrestrial
Planet Finder Foundation Science Program.  
The Lyot Project is also grateful to the Cordelia Corporation, Hilary and Ethel Lipsitz,
the Vincent Astor Fund, Judy Vale and an anonymous donor.
This work has been partially supported by
the National Science Foundation Science and Technology Center for Adaptive Optics, 
managed by the University of California at Santa Cruz under cooperative
agreement AST-9876783.

\bibliographystyle{apj}
\bibliography{ms}

\begin{thebibliography}{44}
\expandafter\ifx\csname natexlab\endcsname\relax\def\natexlab#1{#1}\fi

\bibitem[{{Abramowitz} \& {Stegun}(1970)}]{AbramowitzStegun}
{Abramowitz}, M. \& {Stegun}, I.~A. 1970, {Handbook of Mathematical Functions}
  (Handbook of Mathematical Functions, New York: Dover)

\bibitem[{{Aime} \& {Soummer}(2004)}]{AS04ApJL}
{Aime}, C. \& {Soummer}, R. 2004, \apjl, 612, L85

\bibitem[{{Bloemhof} {et~al.}(2001){Bloemhof}, {Dekany}, {Troy}, \&
  {Oppenheimer}}]{Bloemhof01}
{Bloemhof}, E.~E., {Dekany}, R.~G., {Troy}, M., \& {Oppenheimer}, B.~R. 2001,
  \apjl, 558, L71

\bibitem[{{Bracewell}(1986)}]{Bracewell}
{Bracewell}, R.~N. 1986, {The Fourier Transform and its applications}
  (McGraw-Hill Series in Electrical Engineering, Networks and Systems, New
  York: McGraw-Hill, 2nd rev.ed.)

\bibitem[{{Burrows} {et~al.}(2004){Burrows}, {Sudarsky}, \&
  {Hubeny}}]{Burrows04}
{Burrows}, A., {Sudarsky}, D., \& {Hubeny}, I. 2004, \apj, 609, 407

\bibitem[{{Chanan} \& {Troy}(1999)}]{Chanan98}
{Chanan}, G. \& {Troy}, M. 1999, \ao, 38, 6642

\bibitem[{{Digby} {et~al.}(2006){Digby}, {Hinkley}, {Oppenheimer},
  {Sivaramakrishnan}, {Lloyd}, {Perrin}, {Roberts}, {Soummer}, {Brenner},
  {Makidon}, {Shara}, {Graham}, {Kalas}, {Kuhn}, \& {Newburgh}}]{Digby06}
{Digby}, A.~P., {Hinkley}, S., {Oppenheimer}, B.~R., {Sivaramakrishnan}, A.,
  {Lloyd}, J.~P., {Perrin}, M.~D., {Roberts}, L.~C., J., {Soummer}, R.,
  {Brenner}, D., {Makidon}, R.~B., {Shara}, M., {Graham}, J.~R., {Kalas},
  P.~D., {Kuhn}, J.~R., \& {Newburgh}, L. 2006, \apj, submitted

\bibitem[{Frigo \& Johnson(1997)}]{fftw}
Frigo, M. \& Johnson, S.~G. 1997, in Technical Report MIT-LCS-TR-728
  (Massachusetts Institute of Technology)

\bibitem[{Greenfield {et~al.}(2003)Greenfield, Miller, Hsu, \& White}]{pycon03}
Greenfield, P., Miller, J.~T., Hsu, J.-C., \& White, R.~L. 2003, in PyCon 2003
  Proceedings, ed. S.~Holden

\bibitem[{{Greenfield} {et~al.}(2002){Greenfield}, {Miller}, {Hsu}, \&
  {White}}]{numarray02}
{Greenfield}, P., {Miller}, T., {Hsu}, J.-C., \& {White}, R.~L. 2002, in ASP
  Conf. Ser. 281: Astronomical Data Analysis Software and Systems XI, 140--+

\bibitem[{{Hinkley} {et~al.}(2006){Hinkley}, {Oppenheimer}, {Brenner},
  {Perrin}, {Sivaramakrishnan}, {Roberts}, {Soummer}, \& {Makidon}}]{Hinkley06}
{Hinkley}, S., {Oppenheimer}, B.~R., {Brenner}, D., {Perrin}, M.~D.,
  {Sivaramakrishnan}, A., {Roberts}, L.~C., J., {Soummer}, R., \& {Makidon},
  R.~B. 2006, \apj, submitted

\bibitem[{{Kuchner} \& {Traub}(2002)}]{KT02}
{Kuchner}, M.~J. \& {Traub}, W.~A. 2002, \apj, 570, 900

\bibitem[{{Lane} {et~al.}(1992){Lane}, {Glindemann}, \& {Dainty}}]{Lane92}
{Lane}, R.~G., {Glindemann}, A., \& {Dainty}, J.~C. 1992, Waves Random Media,
  2, 209

\bibitem[{{Lloyd} \& {Sivaramakrishnan}(2005)}]{LS05}
{Lloyd}, J.~P. \& {Sivaramakrishnan}, A. 2005, \apj, 621, 1153

\bibitem[{{Lyot}(1939)}]{Lyot39}
{Lyot}, B. 1939, \mnras, 99, 580

\bibitem[{{Macintosh} {et~al.}(2004){Macintosh}, {Bauman}, {Wilhelmsen Evans},
  {Graham}, {Lockwood}, {Poyneer}, {Dillon}, {Gavel}, {Green}, {Lloyd},
  {Makidon}, {Olivier}, {Palmer}, {Perrin}, {Severson}, {Sheinis},
  {Sivaramakrishnan}, {Sommargren}, {Soummer}, {Troy}, {Wallace}, \&
  {Wishnow}}]{bmac2004SPIE}
{Macintosh}, B.~A., {Bauman}, B., {Wilhelmsen Evans}, J., {Graham}, J.~R.,
  {Lockwood}, C., {Poyneer}, L., {Dillon}, D., {Gavel}, D.~T., {Green}, J.~J.,
  {Lloyd}, J.~P., {Makidon}, R.~B., {Olivier}, S., {Palmer}, D., {Perrin},
  M.~D., {Severson}, S., {Sheinis}, A.~I., {Sivaramakrishnan}, A.,
  {Sommargren}, G., {Soummer}, R., {Troy}, M., {Wallace}, J.~K., \& {Wishnow},
  E. 2004, in Advancements in Adaptive Optics, Proceedings of the SPIE, Volume
  5490, 359--369

\bibitem[{{Makidon} {et~al.}(2005){Makidon}, {Sivaramakrishnan}, {Perrin},
  {Roberts}, {Oppenheimer}, {Soummer}, \& {Graham}}]{Makidon05}
{Makidon}, R.~B., {Sivaramakrishnan}, A., {Perrin}, M.~D., {Roberts}, L.~C.,
  {Oppenheimer}, B.~R., {Soummer}, R., \& {Graham}, J.~R. 2005, \pasp, 117, 831

\bibitem[{Mar\'echal(1947)}]{marechal}
Mar\'echal, A. 1947, Rev. d'Opt, 26, 257

\bibitem[{Marois(2004)}]{Marois04phd}
Marois, C. 2004, PhD thesis, Universit\'e de Montr\'eal, Montr\'eal, Qu\'ebec,
  Canada

\bibitem[{{Marois} {et~al.}(2000){Marois}, {Doyon}, {Racine}, \&
  {Nadeau}}]{Marois00}
{Marois}, C., {Doyon}, R., {Racine}, R., \& {Nadeau}, D. 2000, \pasp, 112, 91

\bibitem[{{Marois} {et~al.}(2006){Marois}, Macintosh, Lafreni\`ere, \&
  Doyon}]{Marois06}
{Marois}, C., Macintosh, B.~A., Lafreni\`ere, D., \& Doyon, R. 2006, \apj,
  submitted

\bibitem[{{Marois} {et~al.}(2004){Marois}, {Racine}, {Doyon}, {Lafreni{\`e}re},
  \& {Nadeau}}]{Marois04}
{Marois}, C., {Racine}, R., {Doyon}, R., {Lafreni{\`e}re}, D., \& {Nadeau}, D.
  2004, \apjl, 615, L61

\bibitem[{{Mouillet} {et~al.}(2004){Mouillet}, {Lagrange}, {Beuzit}, {Moutou},
  {Saisse}, {Ferrari}, {Fusco}, \& {Boccaletti}}]{mouillet2004ASPC}
{Mouillet}, D., {Lagrange}, A.~M., {Beuzit}, J.-L., {Moutou}, C., {Saisse}, M.,
  {Ferrari}, M., {Fusco}, T., \& {Boccaletti}, A. 2004, in ASP Conf. Ser. 321:
  Extrasolar Planets: Today and Tomorrow, 39--+

\bibitem[{{Oppenheimer} {et~al.}(2004){Oppenheimer}, {Digby}, {Newburgh},
  {Brenner}, {Shara}, {Mey}, {Mandeville}, {Makidon}, {Sivaramakrishnan},
  {Soummer}, {Graham}, {Kalas}, {Perrin}, {Roberts}, {Kuhn}, {Whitman}, \&
  {Lloyd}}]{OppenheimerSPIE04}
{Oppenheimer}, B.~R., {Digby}, A.~P., {Newburgh}, L., {Brenner}, D., {Shara},
  M., {Mey}, J., {Mandeville}, C., {Makidon}, R.~B., {Sivaramakrishnan}, A.,
  {Soummer}, R., {Graham}, J.~R., {Kalas}, P., {Perrin}, M.~D., {Roberts},
  L.~C., {Kuhn}, J.~R., {Whitman}, K., \& {Lloyd}, J.~P. 2004, in Advancements
  in Adaptive Optics, Proceedings of the SPIE, Volume 5490, ed. D.~{Bonaccini
  Calia}, B.~L. {Ellerbroek}, \& R.~{Ragazzoni}, 433--442

\bibitem[{{Oppenheimer} {et~al.}(2001){Oppenheimer}, {Golimowski}, {Kulkarni},
  {Matthews}, {Nakajima}, {Creech-Eakman}, \&
  {Durrance}}]{Oppenheimer018pcsurvey}
{Oppenheimer}, B.~R., {Golimowski}, D.~A., {Kulkarni}, S.~R., {Matthews}, K.,
  {Nakajima}, T., {Creech-Eakman}, M., \& {Durrance}, S.~T. 2001, \aj, 121,
  2189

\bibitem[{{Oppenheimer} {et~al.}(1998){Oppenheimer}, {Kulkarni}, {Matthews}, \&
  {van Kerkwijk}}]{Opp98}
{Oppenheimer}, B.~R., {Kulkarni}, S.~R., {Matthews}, K., \& {van Kerkwijk},
  M.~H. 1998, \apj, 502, 932

\bibitem[{{Oppenheimer} {et~al.}(2003{\natexlab{a}}){Oppenheimer}, {Shara},
  {Graham}, {Kalas}, {Lloyd}, {Makidon}, {Sivaramakrishnan}, {Baudoz}, {Kuhn},
  \& {Potter}}]{OppenheimerAMOS03}
{Oppenheimer}, B.~R., {Shara}, M., {Graham}, J.~R., {Kalas}, P., {Lloyd},
  J.~P., {Makidon}, R.~B., {Sivaramakrishnan}, A., {Baudoz}, P., {Kuhn}, J., \&
  {Potter}, D. 2003{\natexlab{a}}, in 2002 AMOS Technical Conference, P. W.
  Kervin, J. L. Africano; eds.

\bibitem[{{Oppenheimer} {et~al.}(2003{\natexlab{b}}){Oppenheimer},
  {Sivaramakrishnan}, \& {Makidon}}]{OSM03}
{Oppenheimer}, B.~R., {Sivaramakrishnan}, A., \& {Makidon}, R.~B.
  2003{\natexlab{b}}, {Imaging Exoplanets: The Role of Small Telescopes} (The
  Future of Small Telescopes In The New Millennium.~Volume III - Science in the
  Shadow of Giants), 155

\bibitem[{{Perrin} {et~al.}(2003){Perrin}, {Sivaramakrishnan}, {Makidon},
  {Oppenheimer}, \& {Graham}}]{Perrin03}
{Perrin}, M.~D., {Sivaramakrishnan}, A., {Makidon}, R.~B., {Oppenheimer},
  B.~R., \& {Graham}, J.~R. 2003, \apj, 596, 702

\bibitem[{{Poyneer} \& {Macintosh}(2004)}]{Poyneer04}
{Poyneer}, L.~M. \& {Macintosh}, B.~A. 2004, Journal of the Optical Society of
  America A, 21, 810

\bibitem[{{Roberts} \& {Neyman}(2002)}]{LRCN02}
{Roberts}, L.~C. \& {Neyman}, C.~R. 2002, \pasp, 114, 1260

\bibitem[{{Roe}(2002)}]{Roe2002}
{Roe}, H.~G. 2002, \pasp, 114, 450

\bibitem[{{Sivaramakrishnan} \& {Hodge}(2001)}]{SivaramakrishnanGSMT01}
{Sivaramakrishnan}, A. \& {Hodge}, P.~E. 2001, in The GSMT Green Book (National
  Optical Astronomical Observatories)

\bibitem[{{Sivaramakrishnan} {et~al.}(2001){Sivaramakrishnan}, {Koresko},
  {Makidon}, {Berkefeld}, \& {Kuchner}}]{Sivaramakrishnan01}
{Sivaramakrishnan}, A., {Koresko}, C.~D., {Makidon}, R.~B., {Berkefeld}, T., \&
  {Kuchner}, M.~J. 2001, \apj, 552, 397

\bibitem[{{Sivaramakrishnan} \& {Lloyd}(2005)}]{SL05}
{Sivaramakrishnan}, A. \& {Lloyd}, J.~P. 2005, \apj, 633, 528

\bibitem[{{Sivaramakrishnan} {et~al.}(2002){Sivaramakrishnan}, {Lloyd},
  {Hodge}, \& {Macintosh}}]{Sivaramakrishnan02}
{Sivaramakrishnan}, A., {Lloyd}, J.~P., {Hodge}, P.~E., \& {Macintosh}, B.~A.
  2002, \apjl, 581, L59

\bibitem[{{Sivaramakrishnan} {et~al.}(2004){Sivaramakrishnan}, {Makidon},
  {Soummer}, {Macintosh}, {Troy}, {Chanan}, {Lloyd}, {Perrin}, {Graham},
  {Poyneer}, \& {Sheinis}}]{Sivaramakrishnan04xaopi}
{Sivaramakrishnan}, A., {Makidon}, R.~B., {Soummer}, R., {Macintosh}, B.~A.,
  {Troy}, M., {Chanan}, G.~A., {Lloyd}, J.~P., {Perrin}, M.~D., {Graham},
  J.~R., {Poyneer}, L., \& {Sheinis}, A.~I. 2004, in Advancements in Adaptive
  Optics, Proceedings of the SPIE, Volume 5490, ed. D.~{Bonaccini Calia}, B.~L.
  {Ellerbroek}, \& R.~{Ragazzoni}, 535--544

\bibitem[{{Sivaramakrishnan} {et~al.}(2005){Sivaramakrishnan}, {Soummer},
  {Sivaramakrishnan}, {Lloyd}, {Oppenheimer}, \& {Makidon}}]{SSSLOM05}
{Sivaramakrishnan}, A., {Soummer}, R., {Sivaramakrishnan}, A.~V., {Lloyd},
  J.~P., {Oppenheimer}, B.~R., \& {Makidon}, R.~B. 2005, \apj, 634, 1416

\bibitem[{{Sivaramakrishnan} \& {Yaitskova}(2005)}]{SY05}
{Sivaramakrishnan}, A. \& {Yaitskova}, N. 2005, \apjl, 626, L65

\bibitem[{{Soummer}(2005)}]{Soummer05}
{Soummer}, R. 2005, \apjl, 618, L161

\bibitem[{{Soummer} {et~al.}(2003){Soummer}, {Aime}, \& {Falloon}}]{SAF03}
{Soummer}, R., {Aime}, C., \& {Falloon}, P.~E. 2003, \aap, 397, 1161

\bibitem[{{Troy} \& {Chanan}(2003)}]{Troy03}
{Troy}, M. \& {Chanan}, G.~A. 2003, in Future Giant Telescopes, Proceedings of
  the SPIE, Volume 4840, ed. J.~R.~P. {Angel} \& R.~{Gilmozzi}, 81--92

\bibitem[{{Yaitskova} \& {Dohlen}(2002)}]{Yaitskova02}
{Yaitskova}, N. \& {Dohlen}, K. 2002, Optical Society of America Journal, 19,
  1274

\bibitem[{{Yaitskova} {et~al.}(2003){Yaitskova}, {Dohlen}, \&
  {Dierickx}}]{Yaitskova03}
{Yaitskova}, N., {Dohlen}, K., \& {Dierickx}, P. 2003, Optical Society of
  America Journal, 20, 1563

\end{thebibliography}

\clearpage

%


\end{document}